\documentclass[traditabstract]{aa}
\usepackage{graphicx}
\usepackage{txfonts}
\usepackage{natbib}
\usepackage{float}

\newcommand{\texi}{$T_{\rm ex}$}
\newcommand{\ncol}{$N_{\rm mol}$}
\newcommand{\cmi}{cm$^{-2}$}
\newcommand{\water}{H$_2$O}
\newcommand{\chis}{$\chi^2$}
\newcommand{\micron}{$\mu$m}

\begin{document}
%

\title{Warm H$_2$O and OH in the disk around the Herbig star HD 163296}


\author{D.\ Fedele\inst{\ref{inst_mpe}}, S.\ Bruderer\inst{\ref{inst_mpe}}, 
  E.F.\ van Dishoeck\inst{\ref{inst_mpe},\ref{inst_leiden}}, G.J.\ Herczeg\inst{\ref{inst_kavli}},
  N.J.\ Evans II\inst{\ref{inst_austin}}, 
  J.\ Bouwman\inst{\ref{inst_mpia}}, Th.\ Henning\inst{\ref{inst_mpia}}, J.\ Green\inst{\ref{inst_austin}}}

\institute{
Max Planck Institut f\"{u}r Extraterrestrische Physik, Giessenbachstrasse 1, 85748 Garching, Germany\label{inst_mpe}; \\
\and
Leiden Observatory, PO Box 9513, 2300 RA Leiden, The Netherlands\label{inst_leiden};\\
\and
Kavli Institute for Astronomy and Astrophysics, Yi He Yuan Lu 5, Beijing, 100871, P.R. China\label{inst_kavli};\\
\and
University of Texas at Austin, Department of Astronomy, 2515 Speedway,
Stop C1400, Austin TX 78712-1205, USA\label{inst_austin};\\
\and
Max Planck Institute for Astronomy, K\"onigstuhl 17, 69117, Heidelberg, Germany\label{inst_mpia}
}

\date{Received May 16 2012; accepted July 17 2012}

\titlerunning{Warm H$_2$O and OH in the HD 163296 disk}
\authorrunning{Fedele et al.}

\offprints{Davide Fedele,\\ \email{fedele@mpe.mpg.de}}

\abstract { 
  We present observations of far-infrared (50-200\,$\mu$m) OH and \water \
  emission of the disk around the Herbig Ae star HD 163296 obtained with 
  Herschel/PACS in the context of the DIGIT key program. In addition to strong 
  [\ion{O}{i}] emission, a number of OH doublets 
  and a few weak highly excited lines of H$_2$O are detected. The presence of 
  warm \water \ in this Herbig disk is confirmed by a line stacking analysis, 
  enabled by the full PACS spectral scan, and by lines seen in {\it Spitzer} data. 
  The line fluxes are analyzed using an LTE slab model including line opacity. 
  The \water \ column density is $10^{14}-10^{15}$\,cm$^{-2}$, and the excitation 
  temperature is 200-300\,K implying warm gas with a density $n > 10^5$\,cm$^{-3}$. 
  For OH we find \ncol\ of $10^{14}-10^{15}$\,cm$^{-2}$ and \texi \ $\sim$ 300-500\,K.
  For both species we find an emitting region of $r \sim 15-20$\,AU from the star.
  We argue that the molecular emission arises from the protoplanetary disk 
  rather than from an outflow. This far-infrared detection of both \water \ 
  and OH contrasts with near- and mid-infrared observations, which have 
  generally found a lack of water in the inner disk around Herbig AeBe stars 
  due to strong photodissociation of \water. 
  Given the similarity in column density and emitting region, OH and H$_2$O 
  emission seems to arise from an upper layer of the disk atmosphere of HD 163296, 
  probing a new reservoir of water. The slightly lower temperature of H$_2$O 
  compared to OH suggests a vertical stratification of the molecular gas with OH 
  located higher and H$_2$O deeper in the disk, consistent with thermo-chemical models.
}


\maketitle

\section{Introduction} \label{sec:intro} 
Water is a key molecule for the chemical and physical evolution of 
protoplanetary disks. Together with O and OH, it forms the main reservoir 
of oxygen. The formation of water ice layers on dust grains may improve their 
sticking behaviour and thereby help the coagulation process towards larger 
particles that ultimately leads to the formation of planetesimals and 
planets. The growth of icy grains is also likely involved in 
the delivery of water to planets. 
Recent observations 
at near- ($\sim$ 3\,$\mu$m) and mid-infrared ($\sim$ 10--30\,$\mu$m) 
wavelengths have revealed the common presence of hot ($T \sim$ 500-1000\,K) 
water and OH vapor in the
atmosphere of T Tauri disks \citep{Salyk08, Carr08, Pontoppidan10,Mandell12}.
In contrast, disks around the more massive Herbig AeBe stars do
not show hot water vapor emission and appear to be depleted in water
molecules \citep{Mandell08, Pontoppidan10, Fedele11}. A likely
explanation is that hot water is photodissociated by the stronger
ultraviolet radiation emitted by the Herbig stars in the regions
of the disks ($<$ a few AU) probed at these wavelengths \citep[e.g.][]{Fedele11}. 
Indeed, in the case of the young eruptive star EX Lupi the \water \ 
emission is variable as a consequence of the changing UV radiation 
field \citep{Banzatti12}.

However, cooler water may survive further
out or deeper into these disks, but that region can only be probed by longer
wavelength data. {\it Herschel} offers the opportunity to search for
these water lines with high sensitivity. Detections of water in disks
with {\it Herschel} have been reported by \citet{Hogerheijde11} using
HIFI and \citet{Riviere12} using PACS, but these observations 
refer only to disks around T Tauri stars.

\noindent
In this letter we report the detection of OH far-infrared emission
lines and the signal of warm H$_2$O toward the Herbig Ae star HD
163296 (A1V) at a distance of $d$ = 118 pc \citep{vanLeeuwen07}. The star is
isolated with no evidence of a stellar companion and is surrounded by
a well-studied disk
\citep[e.g.,][]{Mannings97,Grady01,Isella07}. A bipolar
microjet and a series of Herbig-Haro knots are observed at optical and
UV wavelengths perpendicular to the disk
\citep[e.g.,][]{wassell06}. The disk has recently been modeled by
\citet{Tilling12}, who also report upper limits on selected OH and
H$_2$O lines from PACS data obtained in the GASPS Herschel key program.

\begin{figure*}[htb]
\centering
\includegraphics[width=0.23\hsize]{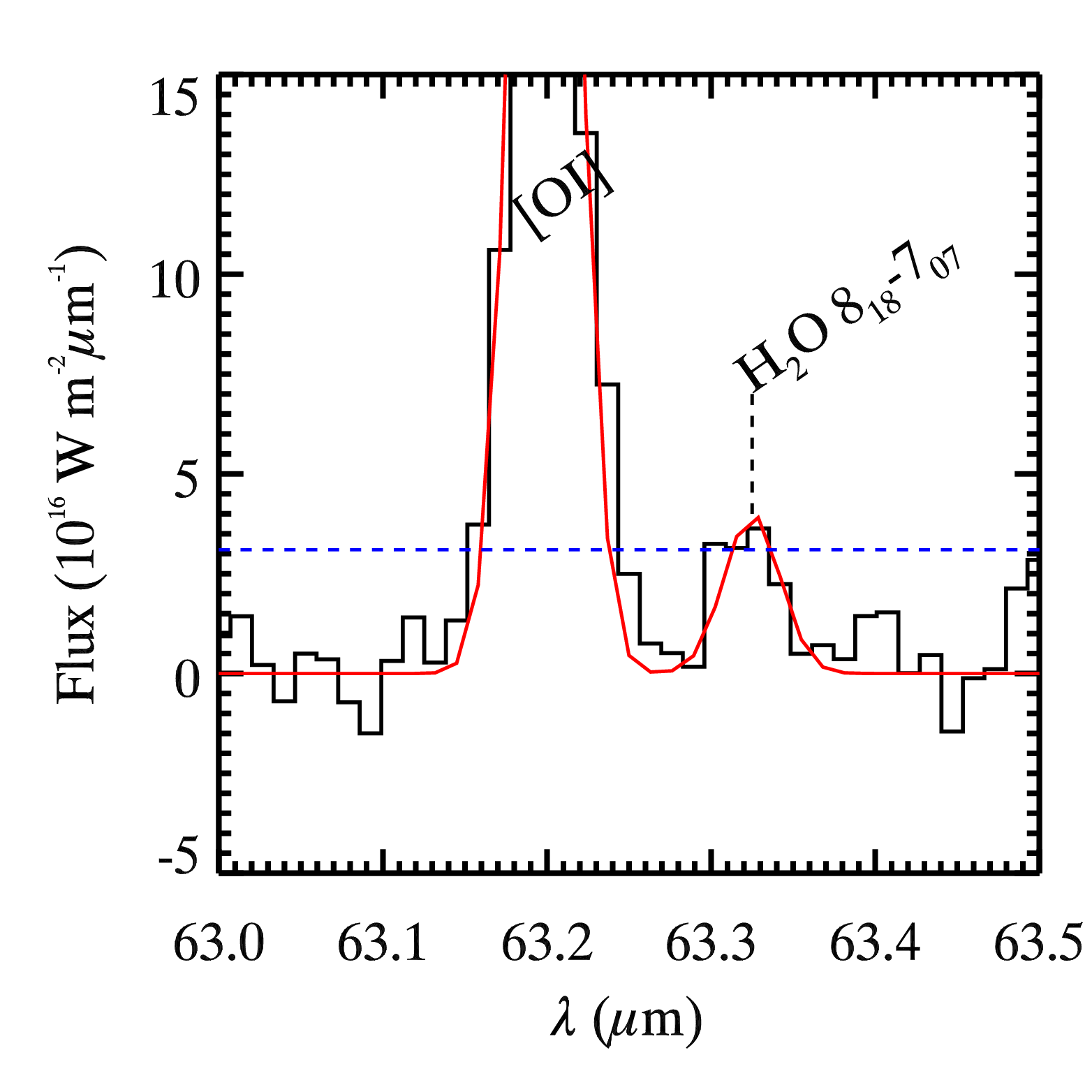}
\includegraphics[width=0.23\hsize]{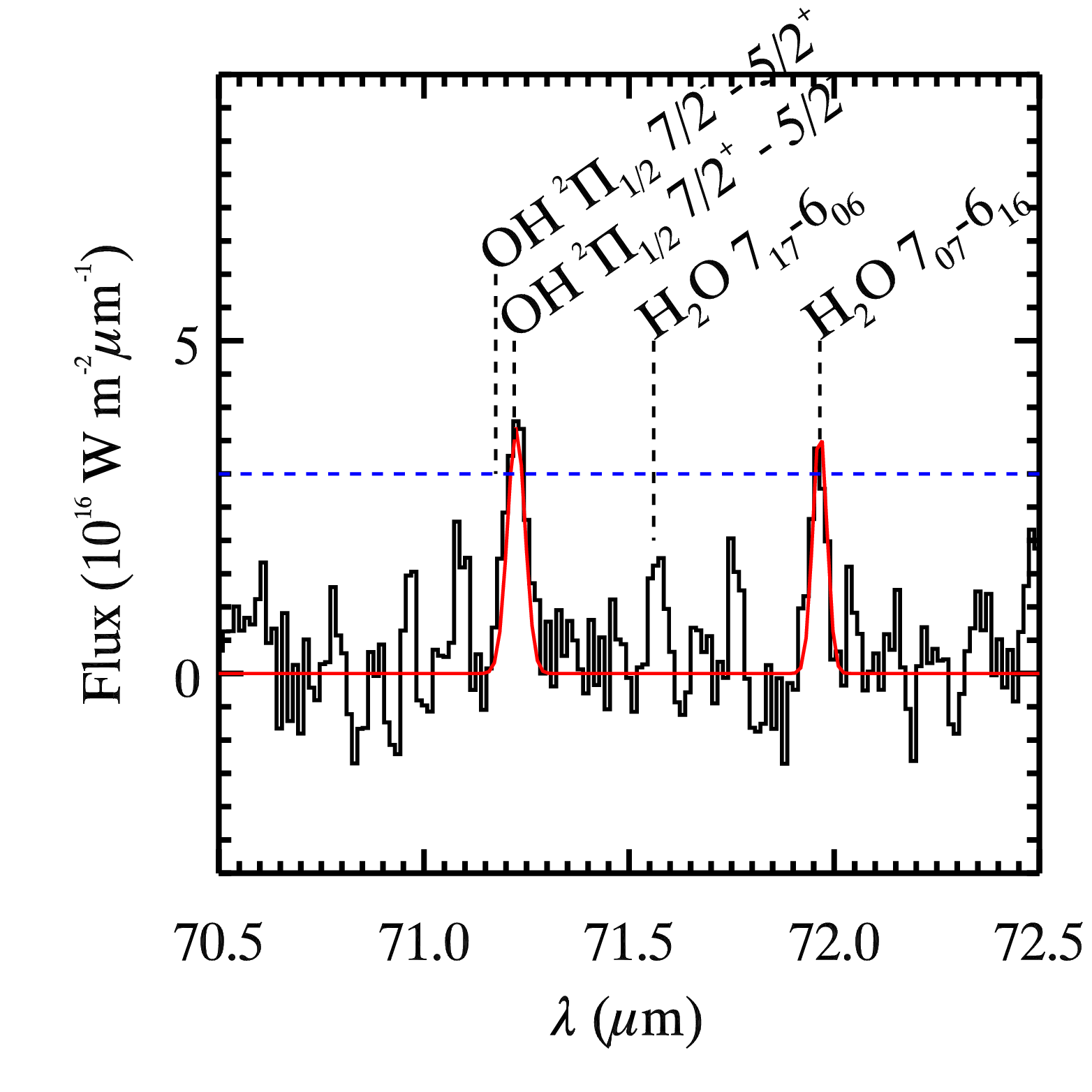}
\includegraphics[width=0.23\hsize]{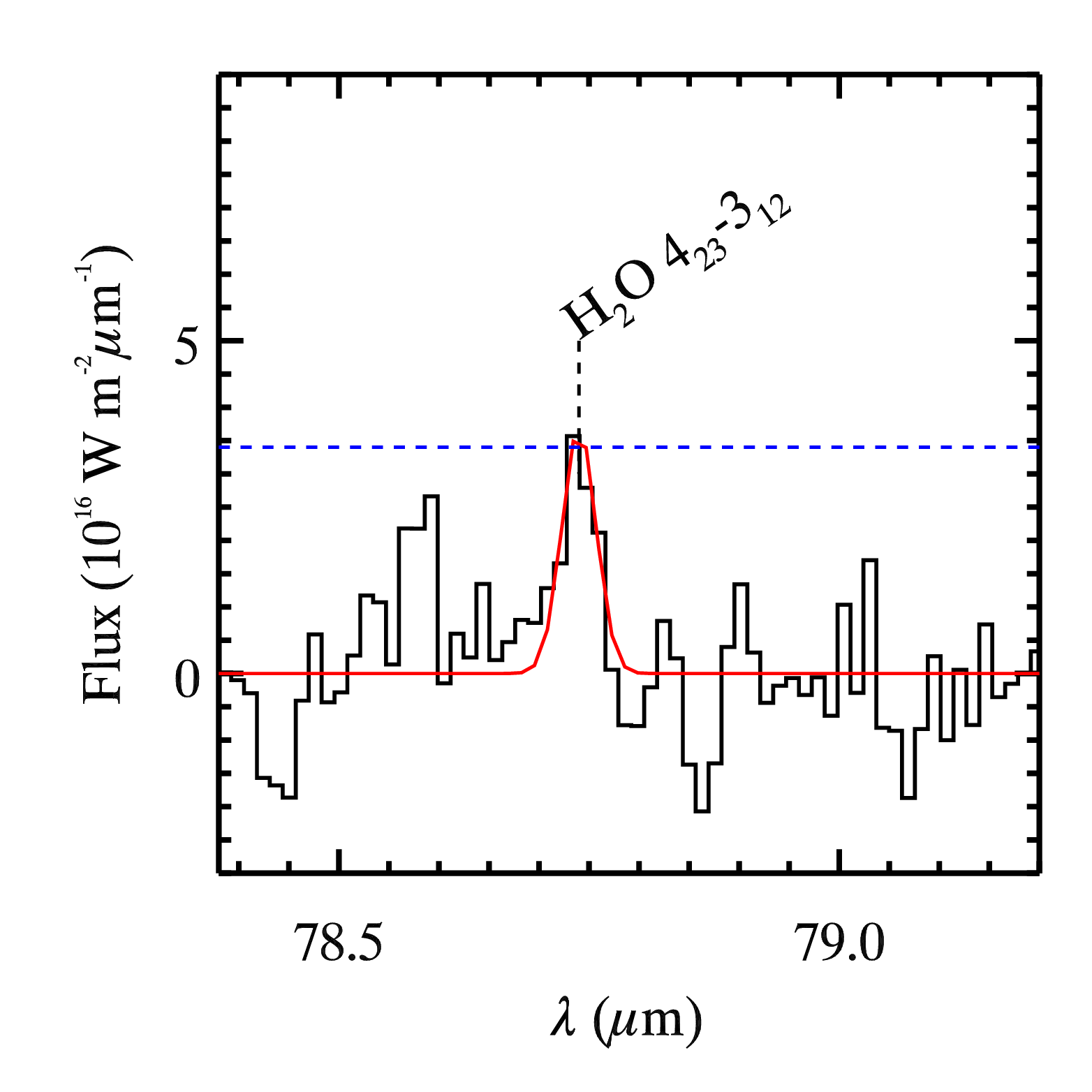}
\includegraphics[width=0.23\hsize]{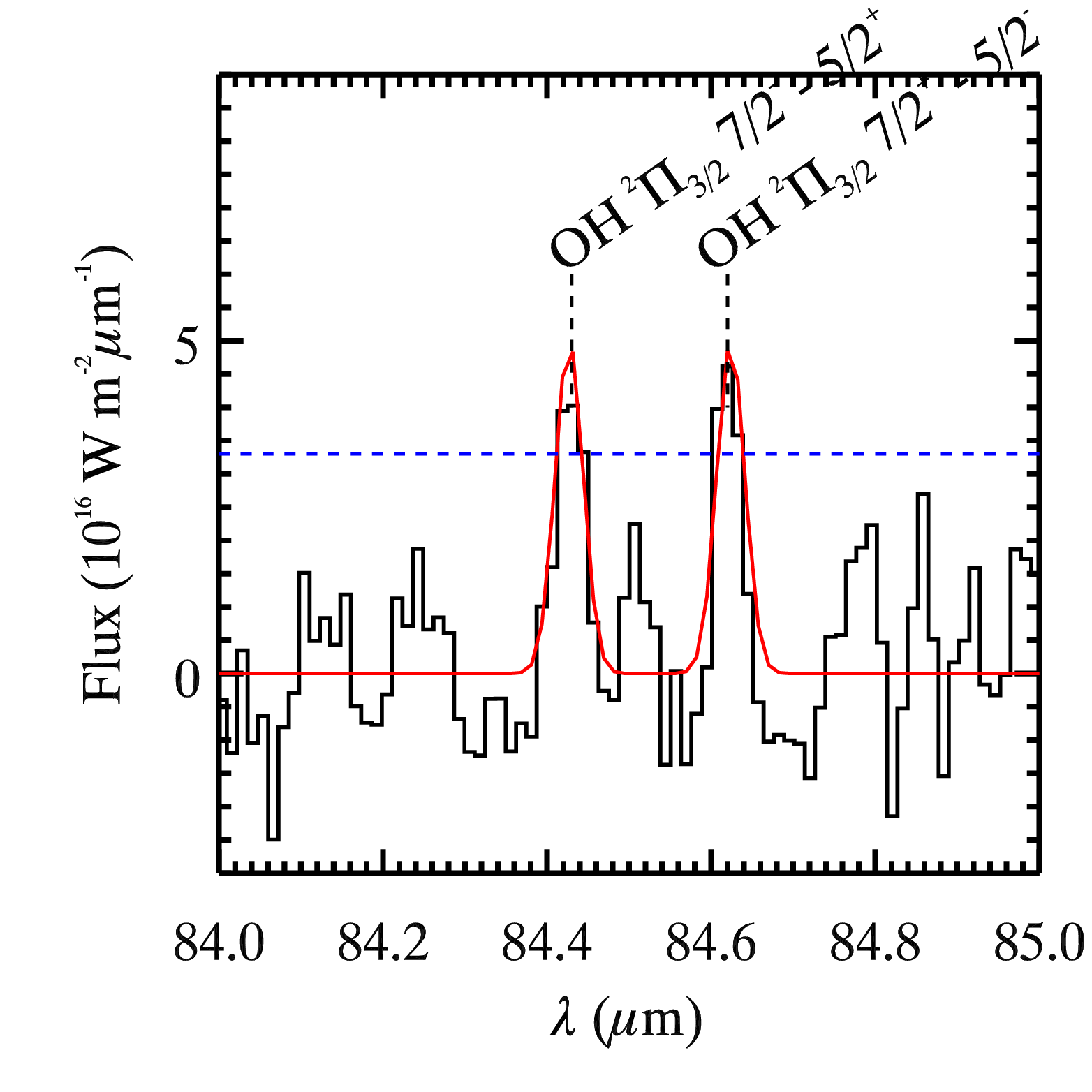}
\caption{PACS spectrum (continuum-subtracted) of selected lines.  The (blue) 
dashed line  indicates the r.m.s. of the baseline multiplied by 3. The 
presented spectrum has been smoothed (smooth width = 2 bins) for clarity. 
The red line is a Gaussian fit to the detected lines.}
\label{fig:spectra}
\end{figure*}

\section{Observations and data reduction} \label{sec:observations} 
HD 163296 was observed on April 03 2011 with the PACS instrument
(\citealt{Poglitsch10}) onboard the {\it Herschel Space
  Observatory} (\citealt{Pilbratt10}) as part of the DIGIT key program
({\it KPOT$\_$nevans$\_$1}, PI: N. Evans). The target was observed in
SED mode covering the wavelength range 50--220\,$\mu$m with $R\sim 1000-3000$ 
(obsid: 1342217819, 1342217829). The observations were carried out in
chopping/nodding mode with a chopping throw of 6\arcmin. The total 
on-source integration time is 6176 seconds for the B2A (51-73\,$\mu$m)
and short R1 (70-105\,$\mu$m) modules and 8360 seconds for the B2B
(70-105\,$\mu$m) and long R1 (140-220\,$\mu$m) modules.  The data have
been reduced with HIPE 8.0.2489 with standard calibration files
from level 0 to level 2.  The two nod positions were reduced
separately (oversampling factor = 3, up-sampling factor = 1 to ensure
that the noise in each spectral point is independent) and
averaged after a flat-field correction.  

\smallskip
The spectrum has been extracted from the central spaxel (9.4$\arcsec$ square) to 
optimize the signal-to-noise ($S/N$) ratio. 
Due to the large pont spread function of the telescope, some flux leaks into the other spaxels
of the PACS array. To recover the absolute flux level, we apply a correction 
factor using the spectrum extracted from the central 9 spaxels (3x3 extraction): this is 
performed by fitting a 3$^{rd}$-order polynomial to two spectra (central spaxel and 3x3), 
the conversion factor is the ratio between the 2 fits. Finally, the spectrum is scaled so that the 
spectrum matches the PACS photometry (from Meeus et al. in preparation) at 70\,\micron \ and 160\,\micron. 


\smallskip
The line flux ($F_{\rm line}$) was measured by fitting a Gaussian function and 
the uncertainty ($\sigma$) is given by the product 
$STD_F \sqrt{\delta \lambda \  FWHM}$\footnote{This formula comes directly from the
error propagation of the sum $\Sigma_i (F_i)$, where $F_i$ is the flux of the $i$-th spectral bin.}, 
where $STD_F$ (W\,m$^{-2}$\,\micron$^{-1}$) is the standard deviation of the (local) spectrum, $\delta \lambda$ is the wavelength spacing of 
the bins (\micron) and $FWHM$ is the full width half maximum of the line 
(\micron).  


\begin{table}
\caption{[\ion{O}{i}] and OH line fluxes.}
\label{tab:lineflux1}
\centering
\begin{tabular}{lrrrrr}
\hline\hline
Transition & $\lambda_{obs}$ & $F_{\rm line}$ & $F_{\rm line}^m$ & $E_u$ & log $A_{ul}$\\
           & (\micron)      & \multicolumn{2}{c}{($10^{-17}$\,W\,m$^{-2}$)} & (K) & (s$^{-1}$)\\
\hline
[\ion{O}{i}]                                 &  63.2 & 19.8$\pm$ 1.2 &     & 228 &  -4.05 \\
$^2\Pi_{1/2}  \, 9/2^+-7/2^-$                  &  55.8 & 5.6 $\pm$ 1.1 & 6.7 & 875 &   0.34 \\
$^2\Pi_{1/2}  \, 9/2^--7/2^+$                  &  55.9 & 6.1 $\pm$ 1.1 & 6.7 & 875 &   0.34 \\
$^2\Pi_{3/2}  \, 9/2^--7/2^+$\tablefootmark{a} &  65.1 & 6.4 $\pm$ 1.2 & 5.9 & 512 &   0.11 \\
$^2\Pi_{3/2}  \, 9/2^+-7/2^-$                  &  65.2 & 5.8 $\pm$ 1.0 & 5.9 & 511 &   0.10 \\
$^2\Pi_{3/2}  \, 7/2^--5/2^+$                  &  84.6 & 2.2 $\pm$ 0.4 & 2.8 & 291 &  -0.28 \\
$^2\Pi_{3/2}  \, 7/2^+-5/2^-$                  &  84.4 & 2.2 $\pm$ 0.4 & 2.8 & 291 &  -0.28 \\
$^2\Pi_{3/2}  \, 5/2^--3/2^+$                  & 119.2 & 1.2 $\pm$ 0.3 & 0.9 & 121 &  -0.86 \\
$^2\Pi_{3/2}  \, 5/2^+-3/2^-$                  & 119.4 & 0.9 $\pm$ 0.3 & 0.9 & 121 &  -0.86 \\
\hline\hline
\end{tabular}
\tablefoot{Column $F_{\rm line}^m$ reports the line flux predicted by the best-fit model. 
\tablefoottext{a}{Blended with o-H$_2$O 6$_{25}$-5$_{14}$}.}
\end{table}

\begin{table}[!ht]
\caption{\water \ line fluxes.}
\label{tab:lineflux2}
\centering
\begin{tabular}{lrrrrr}
\hline\hline
Transition & $\lambda_{obs}$ & $F_{\rm line}$ & $F_{\rm line}^m$ & $E_u$ & log $A_{ul}$\\
           & (\micron)  & \multicolumn{2}{c}{($10^{-17}$\,W\,m$^{-2}$)} & (K) & (s$^{-1}$)\\

\hline
p-H$_2$O \, 4$_{31}$-3$_{22}$ \tablefootmark{a}&  56.31  & 2.7 $\pm$ 1.6 & 2.5 &  552 &  0.16 \\
o-H$_2$O \, 9$_{09}$-8$_{18}$ \tablefootmark{a}&  56.82  & 0.9 $\pm$ 1.6 & 1.5 & 1323 &  0.39 \\
o-H$_2$O \, 8$_{18}$-7$_{07}$                  &  63.32  & 2.0 $\pm$ 0.6 & 2.0 & 1070 &  0.24 \\
o-H$_2$O \, 7$_{07}$-6$_{16}$                  &  71.95  & 2.2 $\pm$ 0.5 & 1.9 &  843 &  0.06 \\
o-H$_2$O \, 4$_{23}$-3$_{12}$                  &  78.74  & 1.8 $\pm$ 0.4 & 1.7 &  432 & -0.32 \\
o-H$_2$O \, 6$_{16}$-5$_{05}$ \tablefootmark{a}&  82.03  & 0.8 $\pm$ 0.8 & 1.5 &  643 &  0.06 \\
p-H$_2$O \, 3$_{22}$-2$_{11}$ \tablefootmark{a}&  89.98  & 0.9 $\pm$ 0.9 & 0.8 &  296 & -0.45 \\
o-H$_2$O \, 2$_{21}$-1$_{10}$ \tablefootmark{a}& 108.07  & 0.7 $\pm$ 0.5 & 0.7 &  194 & -0.59 \\
o-H$_2$O \, 4$_{14}$-3$_{03}$ \tablefootmark{a}& 113.54  & 0.7 $\pm$ 0.4 & 0.6 &  323 & -0.61 \\
\hline\hline
\end{tabular}
\tablefoot{\tablefoottext{a}{Flux integrated over 5 bins centered at the expected line position.}}
\end{table}

\section{Results} \label{sec:results} 
We clearly detect the strong [\ion{O}{i}]\,63\,\micron \ line as well as 5 OH 
far-infrared features above 3$\sigma$ (i.e. having $F_{\rm line}/\sigma > 3$, Table 
\ref{tab:lineflux1}). Spectra of selected lines are shown in Fig. 
\ref{fig:spectra}. The OH lines are readily recognized because of
their doublet pattern; only intra-ladder transitions, which have the
largest Einstein-$A$ coefficients, are found. In the case of the 
$^2\Pi_{1/2} \, 7/2-5/2$ doublet at 71\,\micron \ only one of the two lines is 
detected, although the non-detection of the second line is hardly significant within 
the noise. Asymmetric line intensities of $\Lambda$-doublets are predicted at 
high temperature \citep{Offer92}, but because of the noise this doublet is not considered 
in our further analysis.

\smallskip
Three lines of H$_2$O are detected slightly above 3$\sigma$ (Table \ref{tab:lineflux2}). 
The H$_2$O $8_{18}-7_{07}$ line at 63.32\,$\mu$m is seen not only in our data but also in 
the GASPS spectrum shown by \citet{Tilling12} although they do not claim the 
detection. The H$_2$O $4_{23}-3_{12}$ line at 78.74\,$\mu$m is seen here with a 
flux of 1.8 ($\pm$ 0.4) 10$^{-17}$\,W\,m$^{-2}$ while \citet{Tilling12} report 
only a 3$\sigma$ upper limit of 1.5 10$^{-17}$\,W\,m$^{-2}$.
\citet{Meeus12} recently claim a detection of far-infrared H$_2$O 
emission towards this source based on new GASPS data.
\citet{Pontoppidan10} also provide tentative detections of H$_2$O
lines in the mid-infrared {\it Spitzer} wavelength range. 
Table \ref{tab:lineflux2} summarizes our
fluxes and includes the fluxes measured at the position of 
some (undetected) key H$_2$O lines that are used later in the analysis. 

\smallskip
The detected lines have upper level energies over a wide range with
$E_u/k \sim 120 - 900$ K (OH) and $E_u/k \sim 400 - 1300 $ K (H$_2$O).
Most of the lines are detected in the blue part of the spectrum. 

\begin{figure}[htb]
\centering
\includegraphics[width=0.9\hsize]{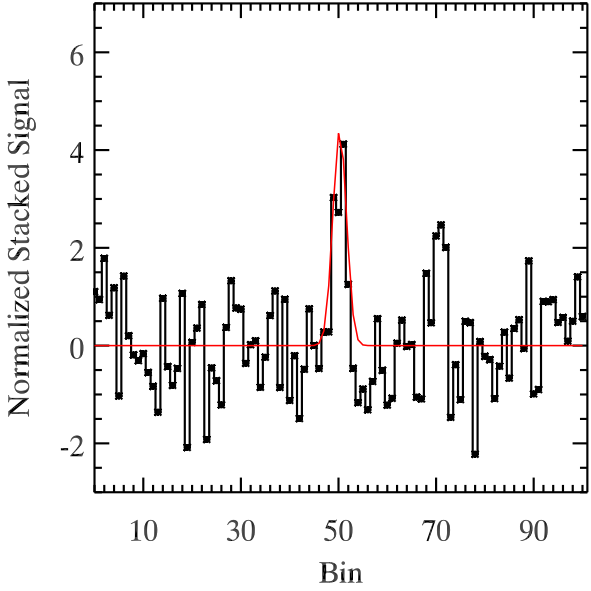}
\caption{
Stacking of 54 H$_2$O lines. Water emission is clearly
detected in the stacked spectrum. The integrated signal is 
$S_{\rm H_2O} = 7\sigma$.}
\label{fig:stacking}
\end{figure}

\subsection{Confirmation of H$_2$O by line stacking}\label{sec:stack}
Since only 3 \water \ lines are marginally detected above 3$\sigma$,
we have used the availability of the full DIGIT PACS spectrum to
confirm the presence of warm water in this disk through a stacking analysis.
Line stacking is commonly done in extra-galactic surveys to detect the
faint emission lines from the outer regions of 
galaxies \citep[e.g.,][]{Schruba11}.
Warm water has many lines spread throughout the far-infrared
wavelength region which can be used for this purpose. 
In this work, we stacked spectra centered at the location of different H$_2$O lines
based on the far-infrared lines detected with PACS towards the protostar 
NGC 1333 IRAS 4B \citep{herczeg12}.  The 95--100 $\mu$m range is excluded 
because of spectral leakage (produced by overlap of grating orders). 
Blended lines are excluded from this analysis and OH and [\ion{O}{i}] lines 
are masked. The remaining number of H$_2$O lines available for the analysis
is 54. The stacked spectrum is the weighted average of 54 spectra, 
each of which is 100 bins wide centered at the position of a water line: 
$F = \frac{\sum_{j=1}^{54} w_j F_j}{\sum_{j=1}^{54} w_j}$
where $F_j$ is the (continuum-subtracted) spectrum centered at
the $j$-th water line and $w_j$ is the weight of the
line.  The weight corresponds to $STD_j^{-1}$, 
where $STD_{j}$ is the standard deviation of the continuum-subtracted 
spectrum $F_j$. The lines are stacked by bin because the spectral resolution
in velocity space varies but is approximately constant in bins. 

\smallskip
The stacked spectrum is shown in Fig. \ref{fig:stacking}. The warm \water
\ signal is clearly detected and centered on the central bin.
The integrated \water \ signal is 7 times its uncertainty. 
The false alarm probability (FAP), i.e. the probability to detect a 
7$\sigma$ signal by stacking random portions of the PACS spectrum, is 
$<$ 0.03\% based on 10,000 randomized tests (see Appendix \ref{sec:fap}).

\begin{figure}[!ht]
\centering
\includegraphics[width=0.9\hsize]{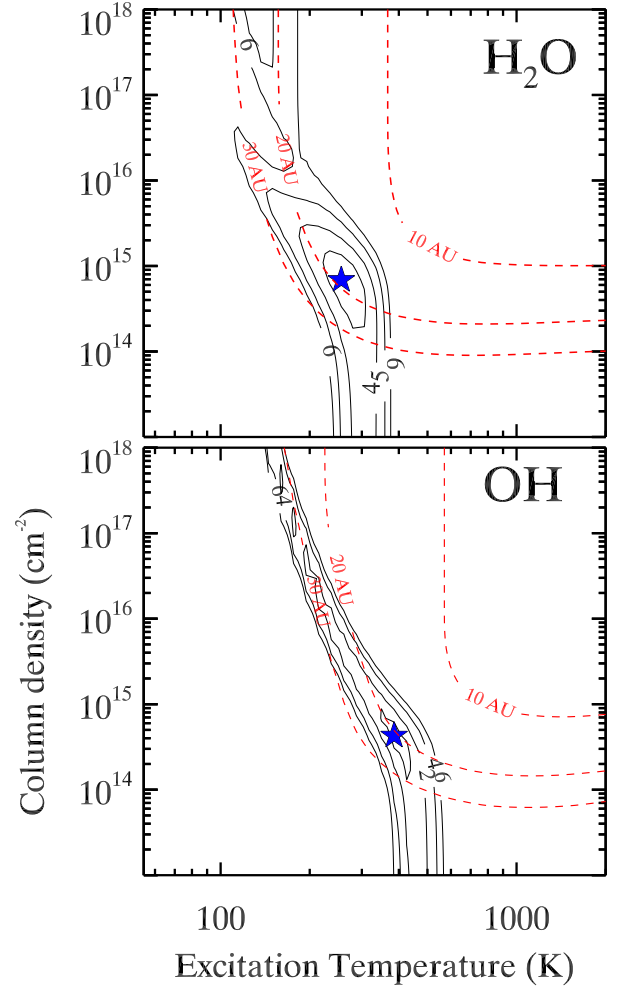}
\caption{Contours of the reduced $\chi^2$ for the slab/LTE model for
  H$_2$O (upper panel) and OH (lower panel). The value of the reduced \chis \ is overplotted. 
  The (red) dashed lines provide 
  the radius of the emitting region (10, 20, 30\,AU). The star indicates
  the location of the minimum $\chi^2$. }
\label{fig:chi2lte}
\end{figure}

\smallskip
This analysis confirms the presence of warm H$_2$O in the PACS
spectrum of HD 163296. Stacking \water \ lines
separately in spectra from the two nod positions also yields $>3
\sigma$ detections. The H$_2$O signal is only detected in the central
spaxel and not in off-source spaxels. These last two
tests exclude the contamination from an extended and/or off-source
emission and confirm that the \water \ lines detected in the PACS
spectrum are associated with HD 163296. 

\begin{figure}
\centering
\includegraphics[width=0.9\hsize]{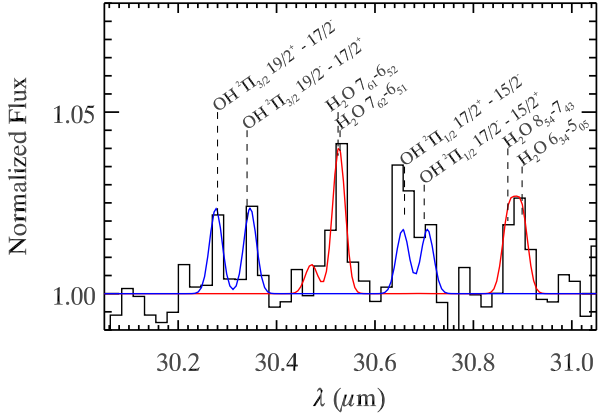}
\caption{{\it Spitzer}/IRS spectrum (continuum-subtracted) of selected lines.
The red and blue lines are the best model for \water \ and OH respectively. 
}
\label{fig:spitzer}
\end{figure}
\section{Analysis}\label{sec:analysis}
In this section we analyze the OH and H$_2$O excitation using a
uniform slab of gas in local thermal equilibrium (LTE) including the
effect of line opacity (see Appendix \ref{sec:rotdiag} for details).
The limited number of lines and their large uncertainties do not warrant
a more sophisticated non-LTE treatment. The analysis is based on the
data in Tables \ref{tab:lineflux1} and \ref{tab:lineflux2}, including
the mid-infrared lines detected with {\it Spitzer}/IRS \citep{Pontoppidan10}. 
Free parameters of the model are the excitation 
temperature \texi \ (K) and the molecular column density \ncol \ (cm$^{-2}$).  
The size of the emitting region, given by its radius $r$, is not a free 
parameter since it can be determined uniquely for every given combination of 
\texi \ and \ncol. Comparison between models and data is done based on reduced
$\chi^2$ values. 

\smallskip
The range of models that yields an acceptable agreement is shown in
Fig. \ref{fig:chi2lte}. Overplotted are contours for the radius $r$.
For H$_2$O, the data are best fitted (1$\sigma$, $p=68.3$\,\%) by models with 
\texi \ $\sim 200-350\,$K and \ncol \ $\sim 10^{14}-10^{16}\,$\cmi \ and 
$r \sim 15-20\,$AU. For OH, the data are best fitted for \ncol \ 
$\sim 10^{14}-10^{15}$\,cm$^{-2}$,  \texi \ $\sim 300-500$\,K and $r \sim 20$\,AU. 
The {\it Spitzer}/IRS spectrum of selected 
lines is shown in Fig. \ref{fig:spitzer} along with the best model. 

\smallskip
Further constraints to the H$_2$O column density and temperature
come from individual line flux ratios. In particular, the ratio between
far- and mid-infrared lines (e.g. $7_{07}-6_{16} / 7_{61}-6_{52}$) constrains
the column density to $> 10^{14}\,$cm$^{-2}$. On the other hand, the ratio
$8_{18}-7_{07} / 8_{08}-7_{17}$ constrain the upper limit of the H$_2$O column density 
to $2 \times 10^{15}\,$cm$^{-2}$. We also note that models with higher column density produce several H$_2$O
lines which are not detected in the PACS and IRS spectra.
The ratio $8_{18}-7_{07} / 9_{09}-8_{18}$ constrains the temperature to $< 300\,$K.

\section{Discussion} \label{sec:discus}
The primary result of this Letter is a detected signal of \water \ toward a 
Herbig star, in addition to OH. The \water \ emitting region is found to be 
15-20\,AU, demonstrating that \water \ can survive the UV radiation further away
from the star while it is likely photodissociated in the inner part of the disk.

\smallskip
Given that a bipolar microjet is known to be associated with HD 163296
\citep{wassell06}, the question arises whether the far-infrared
molecular line emission presented here indeed arises from the disk or
whether it comes from such a jet.  There are several arguments in favor 
of the disk. First, we note that HD 163296 is isolated and
not associated with a molecular cloud.  No evidence of a
molecular outflow has been reported to date \citep[e.g.][]{Bae}. 
Second, the spectrally resolved CO $J=3-2$ line in the sub-millimeter shows the 
characteristic double-peaked profile for gas in Keplerian rotation 
\citep{thi01,Dent05}. 
At much shorter wavelengths
(4.7\,$\mu$m), the CO ro-vibrational emission lines are also
characterized by a double-peaked profile \citep{Salyk11b}. 
Thus, there is no hint of any significant small- or large-scale molecular outflow in these data that could dominate the PACS emission. 
Third, the PACS data show no evidence for extended/off-source
emission beyond the central spaxel, not even for the strong 
[\ion{O}{i}]\,63\,\micron \ line, which places the warm H$_2$O within 500 AU of the central star. 

\smallskip
The inferred OH and H$_2$O excitation temperatures of several hundred
K indicate warm emitting regions. 
The high critical densities of the \water\ lines, $n_c \geq 10^7$ cm$^{-3}$, 
implies that the density of the gas should also be high 
\citep[$n \gtrsim 10^5$\,cm$^{-3}$, e.g.][]{herczeg12}. These conditions
and the arguments above suggest that the OH and H$_2$O emission arises 
from the atmosphere of the disk associated with HD 163296 at radial 
distances $>$10\,AU from the star. 

\smallskip
Assuming that the OH/H$_2$O far-infrared lines are emitted by the
disk, which zone does this emission trace?  Models of the water
chemistry in Herbig disks suggest at least three chemically distinct
zones
\citep[e.g.,][]{Woitke09,Glassgold09,Walsh10,Walsh12,Vasyunin11,Najita11}: (i)
an inner disk water reservoir ($\lesssim$ few AU) with a chemistry
close to LTE; (ii) a cold water belt at large distances ($\gtrsim$50
AU) where gaseous H$_2$O results primarily from photodesorption of
water ice; and (iii) a hot water layer at intermediate distances of
1--30 AU and at medium heights with water formation driven by high
temperature neutral-neutral reactions.  The derived parameters for our
OH and H$_2$O lines are consistent with zone (iii) (see also
\citealt{Tilling12}); zone (i) is probed by the near-infrared data and
zone (ii) can be targeted by HIFI observations of low-$J$ lines.
Thus, the PACS data reveal a new water reservoir in disks.


\section{Conclusions}
We have presented new Herschel/PACS observations of the
disk around the Herbig Ae star HD 163296. We obtain 
detections of far-infrared lines of warm OH and H$_2$O toward a Herbig
star. The presence of warm \water \ is confirmed by a line stacking
analysis (7$\sigma$ detection) enabled by the full PACS
spectral scan. The LTE slab model analysis including optical depth 
effects indicates emission from the intermediate radii of the disk.
Combined with near-infrared and sub-millimeter data, the oxygen chemistry
can now be probed over the entire disk range.

\bibliographystyle{aa}
\bibliography{mybib}

\Online 

\begin{appendix}

\section{Slab model} \label{sec:rotdiag}

For an optically thin line from a point-like source, the flux can be 
written by 

\begin{equation} \label{eq:rotbas}
F_{ul} = d\Omega_{\rm s} \cdot I_{ul} = d\Omega_{\rm s} \frac{h\nu_{ul}}{4\pi} A_{ul} \, N_{\rm mol} \frac{g_u e^{-E_u/k T}}{Q(T)} \,
\end{equation}

with the solid angle of the source $d\Omega_{\rm s}$, the line frequency 
$\nu_{ul}$, the Einstein-A coefficient $A_{ul}$, the molecular column density 
$N_{\rm mol}$, the statistical weight of the upper level $g_u$, the energy of 
the upper level $E_u$ and the partition function $Q(T)$. The solid angle of 
the emitting region can be written as $d\Omega_{\rm s} \equiv \pi r^2/d^2$, with 
the radius of the emitting region $r$ and a distance of $d=118$ pc to HD 163296.
%
%
%
For an optically thick line, the integrated intensity is obtained from 

\begin{equation} \label{eq:iint_full}
I_{ul} = \Delta {\rm v} \frac{\nu_{ul}}{c} B_{\nu_{ul}}(T_{\rm ex}) ( 1 - e^{-\tau_{ul}}) \ .
\end{equation}

with the opacity at the line center 
\begin{equation}
\tau_{ul} = \frac{A_{ul} c^3}{8 \pi \nu_{ul}^3 \Delta {\rm v}}  \left( N_l \frac{g_u}{g_l} - N_u \right) \ .
\end{equation}

The (thermal) width of the lines is assumed to be $\Delta {\rm v} \sim 1$ km s$^{-1}$, 
appropriate for gas at several hundred K and we assume 
a simple square like line profile as e.g. used in the RADEX code 
(\citealt{vdTak07}).

\section{False alarm probability of water detection in the stacked 
  spectrum}\label{sec:fap}
\begin{figure}[htb]
\centering
\includegraphics[width=1.0\hsize]{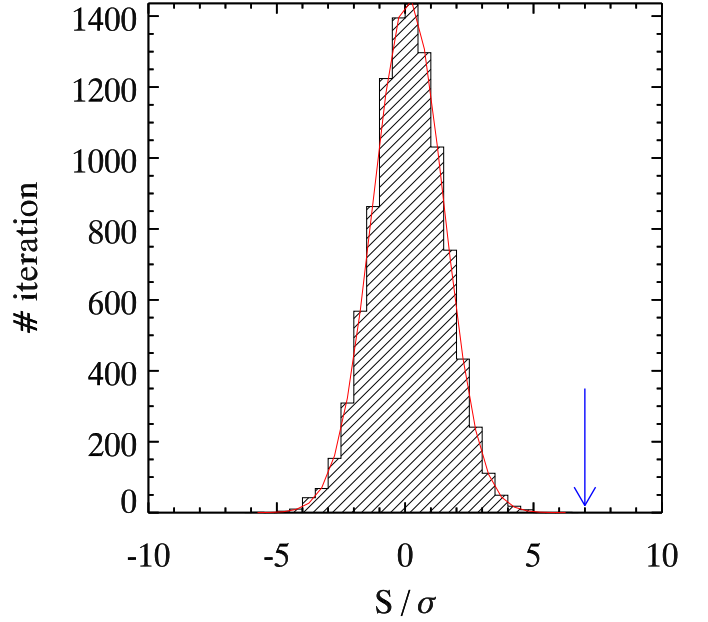}
\caption{Distribution of $S/\sigma$ after 10,000 random stackings of 54 
parts of the PACS spectrum of HD 163296. The red line shows the Gaussian fit. 
The arrow indicates the location of the \water \ signal (Fig. \ref{fig:stacking}).}
\label{fig:fap}
\end{figure}

We performed a simulation to measure the probability to detect a signal with an 
integrated value, $S > 7\sigma$.
This provides the false alarm probability (FAP) of the detection in the stacked 
spectrum. We performed 10,000 random stackings of 54 (equal to the number of water 
lines) parts of the PACS spectrum of HD 163296. After 10,000 iterations we measured 
the distribution of the ratio of the integrated signal over its uncertainty (measured 
as in Sec. \ref{sec:stack}). We mask the bins containing  \water, OH and 
[\ion{O}{i}] emission. Fig. \ref{fig:fap} shows the distribution of $S/\sigma$. 
The distribution is well fitted by a Gauss function (red line, $\tilde{\chi}^2 = 0.03$), 
centered (as expected) at $S/\sigma$ = 0 (i.e. an equal number of 
positive and negative peaks). 
The number of occurrences with $S/\sigma > 7$ is $<$ 3, which corresponds to FAP $<$ 0.03\%
according to Bayesian statistics.

\end{appendix}

\end{document}